\documentclass[cameraready]{Interspeech}
\title{MeCo: One-Step MeanFlow-based Corrector for Multi-Channel Speech Separation}

\author[affiliation={1}]{Dohwan}{Kim}
\author[affiliation={1}, correspondingauthor]{Jung-Woo}{Choi}

\address{
    $^1$ School of Electrical Engineering, KAIST, Daejeon, Republic of Korea
}

\email{dohwan@kaist.ac.kr, jwoo@kaist.ac.kr}


\usepackage{comment}

\usepackage{amsmath,graphicx}
\usepackage{longtable}
\usepackage{amssymb, booktabs}
\usepackage{multirow}
\usepackage{caption}
\usepackage[table]{xcolor}
\usepackage{float}
\usepackage{url}
\usepackage{cite}
\usepackage{graphicx}
\begin{document}

\maketitle

\begin{abstract}
While discriminative models for multi-channel speech separation excel in reference-based metrics, they often exhibit suboptimal human listening quality. To address this, we propose a novel MeanFlow-based one-step generative corrector (MeCo). MeCo learns a conditional average velocity field to map discriminative estimates directly onto the clean speech manifold in a single step. To maximize one-step generation performance, we introduce Data-Space Optimization (DSO). DSO integrates an $\mathbf{x}_r$-loss, which penalizes prediction errors on longer displacement intervals to serve as a generative objective for human listening quality, with an Endpoint SI-SDR loss that directly optimizes terminal signal fidelity. Experiments demonstrate that MeCo achieves state-of-the-art (SOTA) performance with minimal computational overhead, simultaneously achieving superior signal fidelity and human listening quality in both in-domain and out-of-domain scenarios.\footnote{\label{note:code}\scriptsize\url{https://github.com/rlaehghks5/MECO}}
\end{abstract}

\section{Introduction}

Deep discriminative models have significantly advanced multi-channel speech enhancement and separation. Modern architectures \cite{wang2023tf, quan2024spatialnet, kalkhorani2024tf, Lee_2024}, readily adaptable across joint denoising, dereverberation, and speech separation, have achieved saturated performance on reference-based metrics. However, these models are primarily trained to optimize objective metrics such as the Scale-Invariant Signal-to-Distortion Ratio (SI-SDR) \cite{le2019sdr}, which strictly measure signal fidelity and do not perfectly align with human auditory perception. As a result, discriminative models often introduce unnatural speech artifacts \cite{serra2022universal, Richter_2023SGMSE}, a degradation that is clearly captured by reference-free human listening quality estimators such as DNSMOS \cite{reddy2022dnsmos} and UTMOS \cite{saeki2022utmos}.

\vspace{1mm}

To overcome the limitations of discriminative models, generative approaches like diffusion \cite{song2020score} and flow models \cite{lipman2023flow}, have been explored. By learning the underlying distribution of clean speech, they offer better generalization to out-of-domain scenarios and synthesize highly natural audio. Recently, they have been successfully applied to single-channel speech enhancement and separation \cite{lu2022conditional, serra2022universal, Richter_2023SGMSE, lemercier2023storm, scheibler2023DiffSep, liu2023speechflow, dong2025edsep, scheibler2025floss} and multi-channel separation \cite{kimura2025diffcbf, xu2025arraydps}. While recent work on standalone generative models has shown the potential to outperform discriminative baselines in single-channel tasks \cite{scheibler2025floss}, these models are often restricted to constrained environments (e.g., anechoic mixtures without background noise). Furthermore, they suffer from severe inference latency, as multi-channel separators \cite{kimura2025diffcbf, xu2025arraydps} require a large number of reverse steps, making them computationally prohibitive for practical deployment.

Consequently, to efficiently harness the strengths of both approaches in complex, real-world environments, a cascade paradigm has emerged, primarily in single-channel tasks: utilizing a discriminative model for initial separation followed by a generative model for refinement \cite{Sawata_2023, lutati2023separate, wang2024noise, li2025speechrefiner}. This hybrid approach successfully preserves the high signal fidelity of the separator while leveraging the generative prior to repair artifacts and boost generalization. However, most existing refinement models \cite{Sawata_2023, lutati2023separate, li2025speechrefiner} still rely on iterative generation, remaining fundamentally bottlenecked by their inference speed, which limits their feasibility for real-world applications. To address this, Fast-GeCo \cite{wang2024noise} proposed a one-step corrector that outperformed the previous iterative corrector \cite{Sawata_2023} and standalone generative models \cite{scheibler2023DiffSep, liu2023speechflow}. However, Fast-GeCo requires a computationally expensive two-stage training pipeline: pretraining a multi-step diffusion corrector via score matching, then fine-tuning it into a single-step model. Furthermore, to accelerate inference, it heuristically truncates the trajectory at an intermediate time step (e.g., $t=0.5$), thereby introducing a prior mismatch between the terminating forward distribution and the initial reverse distribution. Finally, its one-step fine-tuning relies exclusively on a Scale-Invariant Signal-to-Noise Ratio (SI-SNR) objective. As previously noted, this metric does not align perfectly with human perception, leading to suboptimal results on reference-free human listening quality assessments.

We propose \textbf{MeanFlow-based One-Step Corrector (MeCo)}, a novel generative corrector for one-step multi-channel speech separation without fine-tuning. MeCo is built upon the Mean Flows \cite{geng2025MeanFlow}, which learns the average velocity, capturing the total displacement over a finite interval. By adopting the Mean Flows, which directly map the initial discriminative output ($t=1$) to the clean speech ($t=0$) without heuristic trajectory truncation, MeCo inherently resolves the distribution-mismatch problem. Furthermore, to maximize one-step generation performance, we introduce Data-Space Optimization (DSO), inspired by the analysis of data-space ($\mathbf{x}$-space) objectives \cite{li2026back}. DSO incorporates two complementary objectives: first, an $\mathbf{x}_r$-loss that explicitly penalizes prediction errors on longer displacement intervals; second, an Endpoint SI-SDR loss that incentivizes direct inference from the discriminative output to the terminal state. Specifically, the Endpoint SI-SDR loss ensures high signal fidelity, while the $\mathbf{x}_r$-loss functions as a generative objective, enabling superior human listening quality. To the best of our knowledge, this work is the first to extend one-step generative correctors to multi-channel speech separation. Experimental results demonstrate that MeCo achieves SOTA performance, with high signal fidelity and superior human listening quality on both in-domain and out-of-domain datasets.

\section{Background}
\vspace{-0.5mm}
\subsection{Flow Matching}
\vspace{-0.5mm}

Flow Matching (FM) \cite{lipman2023flow} is a generative framework that learns to construct a flow path between a simple prior distribution $p_0$ and a complex data distribution $p_1$. Formally, given a prior sample $x_0 \sim p_0$ and a data sample $x_1 \sim p_1$, a state $x_t$ along the flow path at time $t \in [0, 1]$ can be explicitly constructed using predefined schedules. In Conditional Flow Matching (CFM) \cite{lipman2023flow}, this state is parameterized by a conditional Gaussian probability path $p_t(x_t \mid x_0, x_1) = \mathcal{N}(x_t \mid \mu_t, \sigma_t^2 \mathbf{I})$ as
\begin{equation}
    x_t = \mu_t + \sigma_t z, \quad z \sim \mathcal{N}(\mathbf{0}, \mathbf{I}).
    \label{eq:xt_path}
\end{equation}

\noindent Here, $\mu_t$ and $\sigma_t$ are predefined mean and standard deviation schedules. For example, a widely adopted Optimal Transport (OT) path defines schedules as linear interpolations over time:
\begin{equation}
    \mu_t = (1-t)x_0 + t x_1, \quad \sigma_t = (1-t)\sigma
\end{equation}
which explicitly constructs a straight-line trajectory between the source and target distributions.

The generative process tracking this path is governed by an Ordinary Differential Equation (ODE) defined by a time-dependent, instantaneous velocity field $v_t$:
\begin{equation}
    dx_t = v_t(x_t) dt.
    \label{eq:fm_ode}
\end{equation}

Because the velocity represents the tangent to the trajectory, the conditional instantaneous target velocity can be analytically derived by taking the time derivative of $x_t$:
\begin{equation}
    v_t(x_t \mid x_0, x_1) = \frac{\sigma_t'}{\sigma_t}(x_t - \mu_t) + \mu_t'
    \label{eq:fm_target_v}
\end{equation}
where $(\cdot)'$ denotes the time derivative. The neural vector field $v_\theta(x_t, t)$ is trained to approximate this target by minimizing the CFM objective:
\begin{equation}
    \mathcal{L}_{\text{CFM}} = \mathbb{E}_{t, x_0, x_1, z} \left[ \left\lVert v_\theta(x_t, t) - v_t(x_t \mid x_0, x_1) \right\rVert^2 \right].
    \label{eq:fm_loss}
\end{equation}

\noindent While CFM provides a computationally efficient training paradigm, sampling from the trained model remains a significant bottleneck. Because the network strictly learns the instantaneous velocity, generating data requires solving the ODE (Eq.~\eqref{eq:fm_ode}) using iterative numerical solvers, such as the Euler method. This reliance on sequential integration steps inherently leads to a high Number of Function Evaluations (NFE) \cite{lipman2023flow, geng2025MeanFlow}, resulting in heavy computational complexity that limits the practical application of standard FM in speech processing tasks \cite{li2025meanflowse, wang2025meanse}.

\vspace{-0.5mm}
\subsection{Mean Flows}
\vspace{-0.5mm}

To overcome the inference bottleneck of standard FM, \textbf{Mean Flows} \cite{geng2025MeanFlow} introduces the concept of the \textbf{average velocity field}. 
Instead of modeling the instantaneous velocity $v_t$, Mean Flows learns to directly predict the average velocity field $u(x_t, r, t)$ over a finite time interval $[r, t]$, where $0 \le r < t \le 1$. 
The average velocity within the interval $[r, t]$ is defined as the time integral of the marginal instantaneous velocity field:
\vspace{-2mm}

\begin{equation}
    u(x_t, r, t) \triangleq \frac{1}{t-r} \int_{r}^{t} v(x_\tau, \tau) d\tau.
        \label{eq:MeanFlow_field}
\end{equation}

While computing this integral during training is intractable, differentiating it with respect to $t$ yields the foundational \textbf{MeanFlow identity}. This identity establishes a tractable relationship between the instantaneous field $v$ and the average field $u$:
\begin{equation}
    u(x_t, r, t) = v(x_t, t) - (t-r) \frac{d}{dt} u(x_t, r, t)
    \label{eq:MeanFlow_identity}
\end{equation}

\noindent where the total derivative is expanded by the chain rule involving a Jacobian-Vector Product (JVP) as
\vspace{-1mm}
\begin{equation}
    \frac{d}{dt} u(x_t, r, t) = v(x_t, t) \partial_x u + \partial_t u.
    \label{eq:MeanFlow_jvp}
\end{equation}
\vspace{-1mm}

To approximate the average field, a neural network $u_\theta(x_t, r, t)$ is trained to estimate the target $u_{\text{tgt}}$ given by
\begin{equation}
    u_{\text{tgt}} = v(x_t, t) - (t-r) \left( v(x_t, t) \partial_x u_\theta + \partial_t u_\theta \right).
    \label{eq:MeanFlow_tgt}
\end{equation}

The model is trained by minimizing the MeanFlow loss $\mathcal{L}_{\text{MF}}$ with a stop-gradient ($\text{sg}$) operator applied to the target:
\begin{equation}
    \mathcal{L}_{\text{MF}} = \mathbb{E} \left[ \left\lVert u_\theta(x_t, r, t) - \text{sg}(u_{\text{tgt}}) \right\rVert^2 \right].
    \label{eq:MeanFlow_loss}
\end{equation}

During inference, the average velocity provides a general displacement rule between any $r < t$:
\vspace{-1mm}
\begin{equation}
    x_r = x_t - (t-r) u_\theta(x_t, r, t).
\end{equation}
\vspace{-1mm}

By evaluating this rule from the initial state ($t=1$) to the terminal state ($r=0$), the model directly transports the data in a single pass, achieving one-step generation at minimal computational cost:
\vspace{-1mm}
\begin{equation}
    x_0 = x_1 - u_\theta(x_1, r=0, t=1).
    \label{eq:MeanFlow_inference}
\end{equation}

\vspace{-3mm}
\section{Method}
\vspace{-1mm}
We introduce MeCo, a one-step generative corrector for multi-channel speech separation. MeCo incorporates a conditional MeanFlow-based architecture (Section \ref{ssec:3.1}) and DSO to maximize one-step generation performance (Section \ref{ssec:3.2}). 
\vspace{-2.5mm}
\subsection{Conditional MeanFlow-based correction}
\label{ssec:3.1}
\vspace{-1mm}
 The proposed corrector operates in the complex Short-Time Fourier Transform (STFT) domain. Let $\mathbf{y} \in \mathbb{C}^{C \times F \times K}$ denote the STFT of a multi-channel noisy mixture with $C$ microphones, $F$ frequency bins, and $K$ time frames, containing $M$ speakers. Let $\mathbf{s}_m \in \mathbb{C}^{F \times K}$ be the corresponding single-channel clean speech for the $m$-th speaker, where $m \in \{1, \dots, M\}$. A discriminative separator processes the mixture to generate an initial, albeit imperfect, estimate $\hat{\mathbf{s}}_m \in \mathbb{C}^{F \times K}$ for each speaker. The corrector refines each speaker's estimate independently by conditioning the generative process on both the multi-channel spatial context $\mathbf{y}$ and the $m$-th speaker's estimate $\hat{\mathbf{s}}_m$. For notational simplicity in the subsequent formulations, we omit the speaker index $m$ and describe the correction process for an arbitrary target speaker, denoting the clean speech and the estimate simply as $\mathbf{s}$ and $\hat{\mathbf{s}}$, respectively.

To transport the distorted discriminative estimate $\hat{\mathbf{s}}$ at $t=1$ to the clean speech manifold $\mathbf{s}$ at $t=0$, we define a conditional Gaussian probability path over $t \in [0, 1]$. During training, $t$ is sampled from $[t_\epsilon, 1]$ to ensure training stability, following the practices in \cite{li2025meanflowse, lee2025flowse}. The time-varying mean $\boldsymbol{\mu}_t$ and standard deviation $\sigma_t$ of the path are explicitly scheduled as:
\vspace{-1.5mm}
\begin{equation}
    \boldsymbol{\mu}_t = (1-t)\mathbf{s} + t\hat{\mathbf{s}}, \quad \sigma_t = (1-t)\sigma_{\text{min}}+t\sigma_{\text{max}}
    \label{eq:path_schedule}
\end{equation}
where $\sigma_{\text{min}}$ and $\sigma_{\text{max}}$ bound the noise variance. An intermediate state $\mathbf{x}_t$ along this trajectory is sampled via:
\vspace{-1.25mm}
\begin{equation}
    \mathbf{x}_t = \boldsymbol{\mu}_t + \sigma_t \mathbf{z}, \quad \mathbf{z} \sim \mathcal{N}(\mathbf{0}, \mathbf{I})
    \label{eq:sampled_state}
\end{equation}
Differentiating the path with respect to time yields the analytical instantaneous velocity target $\mathbf{v}_t$:
\vspace{-0.5mm}

\begin{equation}
    \mathbf{v}_t(\mathbf{x}_t \mid \mathbf{s}, \hat{\mathbf{s}}) = \frac{\sigma_t'}{\sigma_t}(\mathbf{x}_t - \boldsymbol{\mu}_t) + \boldsymbol{\mu}_t'
    \label{eq:instantaneous_v}
\end{equation}
where $\boldsymbol{\mu}_t' = \hat{\mathbf{s}} - \mathbf{s}$ and $\sigma_t' = \sigma_{\text{max}} - \sigma_{\text{min}}$.

By definition, the conditional average velocity field $\mathbf{u}(\mathbf{x}_{t}, r, t \mid \hat{\mathbf{s}})$ represents the integration of the marginal instantaneous velocity $\mathbf{v}(\mathbf{x}_{\tau}, \tau \mid \hat{\mathbf{s}})$ over the interval $[r, t]$, analogous to Eq. (\ref{eq:MeanFlow_field}). To bypass computing this intractable integral, we adapt the MeanFlow identity to directly learn its parameterized estimator $\mathbf{u}_{\theta}(\mathbf{x}_{t},r,t,\mathbf{y},\hat{\mathbf{s}})$, explicitly conditioning the network on the multi-channel mixture $\mathbf{y}$ for additional spatial context. By replacing the marginal field with our analytical on-path instantaneous target $\mathbf{v}_t$, and expanding the total derivative via the chain rule, we formulate the first-order local training target. Following \cite{li2025meanflowse}, we introduce a first-order correction factor $c$ into the formulation to ensure training stability:
\begin{equation}
    \mathbf{u}_{\text{tgt}} = \mathbf{v}_t - c(t-r) \left( \mathbf{v}_t \cdot \nabla_{\mathbf{x}} \mathbf{u}_\theta + \partial_t \mathbf{u}_\theta \right)
    \label{eq:mf_tgt}
\end{equation}
To prevent higher-order backpropagation through the JVP, a stop-gradient ($\text{sg}$) operation is applied to the target. Finally, the model is optimized via the conditional MeanFlow loss:
\begin{equation}
    \mathcal{L}_{\text{MF}} = \mathbb{E} \left[ \left\lVert \mathbf{u}_\theta(\mathbf{x}_t, r, t, \mathbf{y}, \hat{\mathbf{s}}) - \text{sg}(\mathbf{u}_{\text{tgt}}) \right\rVert^2 \right]
    \label{eq:mf_loss}
\end{equation}

\subsection{Data-Space Optimization}
\label{ssec:3.2}
\vspace{-1mm}
To further maximize the performance of one-step generation, we shift the training paradigm from velocity matching to \textbf{Data-Space Optimization (DSO)}. This strategy introduces two complementary training objectives: an $\mathbf{x}_r$-loss that intrinsically scales the matching loss by the displacement distance, and an Endpoint SI-SDR loss that directly optimizes the signal quality.

\textbf{$\mathbf{x}_r$-loss.} 
In the standard MeanFlow objective, the model minimizes the average velocity error. However, minimizing this error alone does not directly reflect the actual deviation in the data space, as it ignores the integration interval. According to the MeanFlow displacement rule, the data at an arbitrary time $r$ is determined by $\mathbf{x}_r = \mathbf{x}_t - \Delta \mathbf{u}$, where $\Delta = t - r$ is the integration interval. To directly optimize the generative trajectory in the data space, we define the predicted data at $r$ as $\hat{\mathbf{x}}_r = \mathbf{x}_t - \Delta \mathbf{u}_\theta$ and the target data at $r$ as $\mathbf{x}_r^{\text{tgt}} = \mathbf{x}_t - \Delta \text{sg}(\mathbf{u}_{\text{tgt}})$. Specifically, we minimize this objective: 
\begin{equation}
    \mathcal{L}_{\mathbf{x}_r} = \mathbb{E} \left[ \left\lVert \mathbf{x}_r^{\text{tgt}} - \hat{\mathbf{x}}_r \right\rVert^2 \right].
\end{equation}

By expanding this objective, we reveal a mathematical relationship with the original velocity loss:
\begin{equation}
    \mathcal{L}_{\mathbf{x}_r} = \scalebox{0.85}{$\displaystyle\mathbb{E} \left[ \left\lVert \left(\mathbf{x}_t - \Delta \text{sg}(\mathbf{u}_{\text{tgt}})\right) - \left(\mathbf{x}_t - \Delta \mathbf{u}_\theta\right) \right\rVert^2 \right] = \Delta^2 \mathcal{L}_{\text{MF}}.$}
\end{equation}

This equation demonstrates that optimizing $\mathbf{x}_r$-loss inherently imposes a time-dependent $\Delta^2$ weighting on the MeanFlow loss. Intuitively, even a small velocity error can lead to a significant displacement if the integration interval $\Delta$ is large. The inherent $\Delta^2$ scaling addresses this by imposing a larger penalty on longer displacement intervals. This transformation is exceptionally advantageous for one-step generation, where $\Delta \approx 1$, as the loss strictly reduces to the direct reconstruction error with respect to the clean speech. 

\textbf{Endpoint SI-SDR Loss.}
While the $\mathbf{x}_r$-loss minimizes the $L_2$ error for any arbitrary $r$, it does not explicitly guarantee the fidelity of the final generated audio. To bridge this gap, we simulate the actual inference process during training by directly estimating the data at the endpoint. Let $t=1$ be the start time corresponding to the initial estimate $\hat{\mathbf{s}}$, and $r=t_\epsilon$ be a terminal time. We directly sample the one-step reconstructed speech $\hat{\mathbf{s}}_{t_\epsilon}$ at the endpoint using the current model prediction:
\begin{equation}
    \hat{\mathbf{s}}_{t_\epsilon} = \mathbf{x}_1 - (1-t_\epsilon) \mathbf{u}_\theta(\mathbf{x}_1, r=t_\epsilon, t=1, \mathbf{y}, \hat{\mathbf{s}}).
\end{equation}
We minimize the negative SI-SDR between this sampled one-step endpoint $\hat{\mathbf{s}}_{t_{\epsilon}}$ and the ground-truth clean speech $\mathbf{s}$:
\begin{equation}
    \mathcal{L}_{\text{SI-SDR}} = - 10 \log_{10} \frac{\left\lVert \alpha \mathbf{s} \right\rVert^2}{\left\lVert \hat{\mathbf{s}}_{t_\epsilon} - \alpha \mathbf{s} \right\rVert^2},
    \label{eq:onestep_sisnr}
\end{equation}
where $\alpha$ is the optimal scaling factor. This forces the model to prioritize velocity fields that yield higher auditory quality when predicting the target speech in a single step. The final training objective combines the $\mathbf{x}_r$-loss and the Endpoint SI-SDR loss:
\begin{equation}
    \mathcal{L}_{\text{MeCo}} = \mathcal{L}_{\mathbf{x}_r} + \mathcal{L}_{\text{SI-SDR}}.
\end{equation}

\vspace{-1mm}
\section{Experiments}
\vspace{-1mm}

\begin{table*}[htbp]
\centering
\caption{Performance comparison on WSJ0 + WHAM! and Librispeech + DEMAND.}
\vspace{-2mm}
\label{tab:performance_comparison}
\setlength{\tabcolsep}{3pt}
\resizebox{\textwidth}{!}{
\begin{tabular}{l | c | c | c | cccccc cccccc}
\toprule
\toprule
\multirow{2}{*}{\textbf{Model}} & \multirow{2}{*}{\textbf{NFE}} & \multirow{2}{*}{\textbf{RTF}} & \multirow{2}{*}{\textbf{Type}} & \multicolumn{6}{c}{\textbf{WSJ0 + WHAM!}} & \multicolumn{6}{c}{\textbf{Librispeech + DEMAND}} \\
\cmidrule(lr){5-10} \cmidrule(lr){11-16}
& & & & $\text{PESQ}^\uparrow$ & $\text{ESTOI}^\uparrow$ & $\text{SI-SDR}^\uparrow$ & $\text{DNSMOS}^\uparrow$ & $\text{UTMOS}^\uparrow$ & $\text{NISQA}^\uparrow$ & $\text{PESQ}^\uparrow$ & $\text{ESTOI}^\uparrow$ & $\text{SI-SDR}^\uparrow$ & $\text{DNSMOS}^\uparrow$ & $\text{UTMOS}^\uparrow$ & $\text{NISQA}^\uparrow$ \\
\midrule
DeFTAN2 & 1 & 0.0155 & D & 1.88 & 0.75 & 9.31 & 2.94 & 3.12 & 3.92 & \textbf{1.78} & 0.71 & 4.96 & 2.88 & 2.90 & 3.62 \\
\cmidrule(lr){1-16}
+ Fast-GeCo (A) & \multirow{3}{*}{+1} & \multirow{3}{*}{+0.0068} & \multirow{3}{*}{D+G} & \textbf{1.96} & 0.79 & 9.81 & 3.11 & 3.51 & 4.11 & 1.75 & 0.72 & 5.10 & 3.08 & 3.22 & 4.00 \\
+ MeanFlow (B) & & & & 1.78 & 0.77 & 10.01 & 3.04 & 3.63 & 4.43 & 1.66 & 0.71 & 5.18 & 3.04 & 3.34 & 4.26 \\
+ MeCo (C) & & & & 1.93 & \textbf{0.80} & \textbf{10.08} & \textbf{3.19} & \textbf{3.70} & \textbf{4.50} & 1.75 & \textbf{0.73} & \textbf{5.19} & \textbf{3.17} & \textbf{3.41} & \textbf{4.38} \\
\midrule
\midrule
SpatialNet & 1 & 0.0078 & D & 1.87 & 0.73 & 8.77 & 2.82 & 2.91 & 3.58 & \textbf{2.15} & 0.80 & 10.00 & 2.87 & 3.15 & 3.76 \\
\cmidrule(lr){1-16}
+ A & \multirow{3}{*}{+1} & \multirow{3}{*}{+0.0068} & \multirow{3}{*}{D+G} & \textbf{1.98} & 0.79 & 9.50 & 3.07 & 3.52 & 3.97 & 2.13 & 0.83 & 10.18 & 3.09 & 3.54 & 4.09 \\
+ B & & & & 1.77 & 0.77 & 9.78 & 3.05 & 3.59 & 4.37 & 1.98 & 0.82 & \textbf{10.43} & 3.07 & 3.65 & 4.41 \\
+ C & & & & 1.95 & \textbf{0.80} & \textbf{9.88} & \textbf{3.18} & \textbf{3.65} & \textbf{4.41} & 2.13 & \textbf{0.84} & 10.41 & \textbf{3.22} & \textbf{3.75} & \textbf{4.52} \\
\midrule
\midrule
CrossNet & 1 & 0.0065 & D & 1.81 & 0.72 & 8.29 & 2.79 & 2.81 & 3.52 & \textbf{2.05} & 0.77 & 8.96 & 2.82 & 3.00 & 3.68 \\
\cmidrule(lr){1-16}
+ A & \multirow{3}{*}{+1} & \multirow{3}{*}{+0.0068} & \multirow{3}{*}{D+G} & \textbf{1.89} & 0.77 & 8.92 & 3.05 & 3.34 & 3.95 & 2.02 & 0.80 & 9.18 & 3.05 & 3.41 & 4.00 \\
+ B & & & & 1.71 & 0.75 & 9.14 & 2.98 & 3.51 & 4.35 & 1.89 & 0.80 & \textbf{9.39} & 3.03 & 3.56 & 4.35 \\
+ C & & & & 1.86 & \textbf{0.78} & \textbf{9.22} & \textbf{3.15} & \textbf{3.56} & \textbf{4.41} & 2.02 & \textbf{0.82} & 9.38 & \textbf{3.16} & \textbf{3.62} & \textbf{4.45} \\
\bottomrule
\bottomrule
\end{tabular}
}
\end{table*}

\begin{table*}[htbp]
\centering
\begin{minipage}[t]{0.48\textwidth}
    \centering
    \vspace{-1.25mm}
    \caption{Ablation study on DSO.}
    \vspace{-2mm}
    \label{tab:ablation_dso}
    \setlength{\tabcolsep}{0.5pt}
    \resizebox{\columnwidth}{!}{
    \begin{tabular}{l | cccc cccc}
    \toprule
    \multirow{2}{*}{\textbf{Model}} & \multicolumn{4}{c}{\textbf{WSJ0 + WHAM!}} & \multicolumn{4}{c}{\textbf{Librispeech + DEMAND}} \\
    \cmidrule(lr){2-5} \cmidrule(lr){6-9}
& $\text{PESQ}^\uparrow$ & $\text{SI-SDR}^\uparrow$ & $\text{DNSMOS}^\uparrow$ & $\text{UTMOS}^\uparrow$ & $\text{PESQ}^\uparrow$ & $\text{SI-SDR}^\uparrow$ & $\text{DNSMOS}^\uparrow$ & $\text{UTMOS}^\uparrow$ \\
    \midrule
    DeFTAN2 & 1.88 & 9.31 & 2.94 & 3.12 & \textbf{1.78} & 4.96 & 2.88 & 2.90 \\
    \midrule
    + MeanFlow & 1.78 & 10.01 & 3.04 & 3.63 & 1.66 & 5.18 & 3.04 & 3.34 \\
    \quad w/ $\mathbf{x}_r$-loss & 1.79 & 10.07 & 3.07 & 3.65 & 1.67 & 5.22 & 3.07 & 3.37 \\
    \quad w/ EP SI-SDR loss & 1.92 & \textbf{10.14} & 3.17 & 3.67 & 1.74 & \textbf{5.26} & 3.15 & 3.37 \\
    + MeCo (Both) & \textbf{1.93} & 10.08 & \textbf{3.19} & \textbf{3.70} & 1.75 & 5.19 & \textbf{3.17} & \textbf{3.41} \\
    \bottomrule
    \end{tabular}
    }
\end{minipage}
\hfill
\begin{minipage}[t]{0.48\textwidth}
    \centering
    \vspace{-1.25mm}
    \caption{Performance comparison on out-of-domain languages.}
    \vspace{-2mm}
    \label{tab:unseen_languages}
    \setlength{\tabcolsep}{3pt}
    \resizebox{\columnwidth}{!}{
    \begin{tabular}{l | c | cccccc}
    \toprule
    \multirow{2}{*}{\textbf{Model}} & \multirow{2}{*}{\textbf{Type}} & \multicolumn{5}{c}{\textbf{Low-resource languages}\cite{sodimana2018multilingual} \textbf{+ DEMAND} } \\
    \cmidrule(lr){3-8}
    &  & $\text{PESQ}^\uparrow$ & $\text{ESTOI}^\uparrow$ & $\text{SI-SDR}^\uparrow$ & $\text{DNSMOS}^\uparrow$ & $\text{UTMOS}^\uparrow$ & $\text{NISQA}^\uparrow$ \\
    \midrule
    DeFTAN2 & D & \textbf{1.74} & 0.73 & 4.87 & 2.84 & 2.36 & 3.68 \\
    \midrule
    + Fast-GeCo & D+G & 1.70 & 0.75 & 5.06 & 3.00 & 2.64 & 3.96 \\
    + MeanFlow  & D+G & 1.64 & 0.74 & 5.06 & 2.99 & 2.75 & 4.33\\
    + MeCo & D+G & \textbf{1.74} & \textbf{0.76} & \textbf{5.08} & \textbf{3.11} & \textbf{2.82} & \textbf{4.38}\\
    \bottomrule
    \end{tabular}
    }
\end{minipage}
\vspace{-5mm}
\end{table*}

\subsection{Datasets}
\vspace{-1mm}
To evaluate the proposed MeCo, we constructed multi-channel noisy and reverberant datasets. For the in-domain training and test sets, we used clean speech from the WSJ0 corpus mixed with noise from WHAM! \cite{wichern2019wham}. To assess the model's generalization capabilities, we constructed two separate out-of-domain evaluation sets. The first dataset, representing an unseen corpus and noise distribution, was constructed by mixing Librispeech \cite{panayotov2015librispeech} with DEMAND \cite{thiemann2013DEMAND} noise. The second dataset, covering unseen languages, comprised six low-resource languages \cite{sodimana2018multilingual} recorded in quiet studio environments and mixed with DEMAND noise \cite{thiemann2013DEMAND}. All audio data were sampled at 16 kHz and segmented into 4-second segments during training.

The simulations were conducted using gpuRIR \cite{diaz2021gpurir}. We simulated a 4-channel circular microphone array (0.05 m radius) in cuboid rooms with width and depth uniformly sampled from $[5.0, 8.0]$ m, height from $[3.0, 4.0]$ m, and RT60 from $[0.2, 0.4]$ s. The microphone array and speakers were randomly positioned at heights of $[1.0, 1.5]$ m and $[1.5, 2.0]$ m, with minimum wall margins of 1.0 m and 0.5 m, respectively. The source-to-array distance ranged between $[0.75, 2.0]$ m. Clean speech and noise were mixed at uniformly distributed SNRs in $[-10, 10]$ dB for single-speaker and in $[10, 20]$ dB for multi-speaker (2- and 3-speaker) mixtures. All models were trained and evaluated blindly, without knowledge of the number of active speakers.

\vspace{-1.25mm}
\subsection{Implementation details}
\vspace{-1mm}
Discriminative and generative models use STFT $n_{\text{fft}}$/hop length of 512/256 and 510/64, respectively. Generative correctors shared the NCSN++ backbone as in \cite{Richter_2023SGMSE}, predicting either the instantaneous velocity field (Fast-GeCo) or the average velocity field $\mathbf{u}_\theta$ (MeanFlow and MeCo). Since no multi-channel one-step correctors currently exist, we adapted the SOTA single-channel one-step corrector, Fast-GeCo \cite{wang2024noise}, as our primary generative baseline. To enable multi-channel processing, all generative networks were conditioned by channel-wise concatenating the complex STFTs of the mixture $\mathbf{y}$ and the discriminative estimate $\hat{\mathbf{s}}$. MeanFlow models integrated a Gaussian Fourier projection and MLP to embed the integration interval $d = t - r$ alongside the standard $t$. Following \cite{li2025meanflowse}, we set $\sigma_{\text{min}}=0.0$, $\sigma_{\text{max}}=0.487$, $c=0.5$, and $t_\epsilon=0.03$. As discriminative separators to provide initial estimates, we employed lightweight versions of three SOTA models: a modified DeFTAN2-base \cite{Lee_2024}, SpatialNet-small \cite{quan2024spatialnet}, and a scaled-down CrossNet \cite{kalkhorani2024tf}. Detailed configurations for each model are available in our public repository.\footref{note:code} 

All models were trained using the Adam \cite{kingma2014adam} optimizer, with the learning rate and batch size set to 1e-3 and 8 for discriminative models, and 1e-4 and 4 for generative correctors, respectively. For generative correctors, we utilized an exponential moving average of 0.999 and gradient clipping at 1.0. Discriminative models were trained for 150 epochs using the negative SA-SDR \cite{von2022sasdr} loss. MeanFlow and MeCo were trained for 100 epochs. In contrast, Fast-GeCo required 100 epochs of score matching for the teacher model, followed by an additional 50 epochs of distillation using only the SI-SDR loss. The best checkpoints were selected based on the highest validation SA-SDR for discriminative models and SI-SDR for generative correctors. To evaluate cross-model generalizability, correctors were trained solely on the output of frozen DeFTAN2 and then evaluated directly on SpatialNet and CrossNet without further training. We report reference-based metrics (PESQ \cite{rix2001pesq}, ESTOI \cite{jensen2016estoi}, SI-SDR) alongside reference-free human listening quality assessments (DNSMOS, UTMOS, NISQA \cite{mittag2021nisqa}). Real-time factor (RTF) was measured on a single RTX 4090 GPU.

\vspace{-1.75mm}
\subsection{Experiment results}
\vspace{-0.5mm}
\subsubsection{Performance on in-domain and out-of-domain}
\vspace{-1mm}

To ensure suitability for practical deployment, our experiments focus exclusively on one-step generative correctors, which introduce a minimal computational overhead of only 1~NFE and an increase of 0.0068 in RTF. As shown in Table~1, MeCo significantly outperforms the evaluated discriminative baselines, the generative corrector Fast-GeCo and MeanFlow across in-domain (WSJ0+WHAM!) and out-of-domain (Librispeech+DEMAND) datasets in almost all objective and subjective metrics. Fast-GeCo relies exclusively on an SI-SNR objective during fine-tuning without a generative loss. Consequently, while it can occasionally surpass MeCo on specific reference-based metrics, like in-domain PESQ, it exhibits lower performance on other reference-based metrics. In particular, it demonstrates significantly inferior performance on reference-free human listening assessments. In contrast, MeCo's DSO integrates a generative $\mathbf{x}_r$-loss to preserve the natural data distribution with an Endpoint SI-SDR Loss for signal fidelity. This unified approach enables MeCo to achieve optimal perceptual quality while preserving deterministic signal enhancement.

\vspace{-2mm}
\subsubsection{Ablation study on DSO}
\vspace{-1mm}
Table 2 details the ablation study on the components of DSO. Adding the $\mathbf{x}_r$-loss and Endpoint SI-SDR loss individually yields incremental improvements over the standard MeanFlow, but combining both in MeCo achieves the highest performance. This is because the $\mathbf{x}_r$-loss imposes a strict penalty on longer displacement intervals during one-step generation, while the Endpoint SI-SDR loss directly optimizes the signal fidelity at the terminal state by the one-step inference.

\vspace{-2mm}
\subsubsection{Generalization to out-of-domain languages}
\vspace{-1mm}
Table 3 evaluates generalization on an out-of-domain dataset comprising six low-resource languages. While DeFTAN2 degrades on unseen languages, MeCo achieves the best performance on both reference-based metrics and reference-free human listening quality assessments. It is notable that Fast-GeCo's generalization performance is constrained because its SI-SNR-only fine-tuning strips away the generalizable generative prior. By learning the underlying distribution of clean speech through the DSO, rather than purely deterministic mappings, MeCo successfully synthesizes highly natural audio and generalizes effectively even to completely unfamiliar linguistic environments.

\vspace{-1.25mm}
\section{Conclusion}
\vspace{-0.75mm}
We proposed MeCo, the first one-step generative corrector for multi-channel speech separation. By leveraging Mean Flows, MeCo effectively maps discriminative estimates directly onto the clean speech manifold in a single step. To maximize one-step generation performance, we introduced DSO, which incorporates an $\mathbf{x}_r$-loss and an Endpoint SI-SDR loss. Experimental results demonstrate that MeCo achieves SOTA performance, excelling in both reference-based metrics and reference-free human listening assessments, with robust generalization across in-domain and out-of-domain datasets. A current limitation is that MeCo's independent speaker refinement relies on channel-wise concatenation. Future work will address this by exploring explicit spatial modeling and joint multi-speaker correction to enhance performance in complex acoustic scenes.

\section{Acknowledgements}
This work was supported by the National Research Foundation of Korea (NRF) grant (No. RS-2024-00337945), STEAM research grant (No. RS-2024-00464269) funded by the Ministry of Science and ICT of Korea government (MSIT), and the BK21 FOUR program through the NRF grant funded by the Ministry of Education of Korea government (MOE).

\section{Generative AI Use Disclosure}
Generative AI tools were used to edit and polish the manuscript, improving readability and refining the experimental code.
\bibliographystyle{IEEEtran}
\bibliography{mybib}

\end{document}